\documentclass[twocolumn,prd,showpacs,preprintnumbers,amsmath,amssymb]{revtex4-1}

\usepackage{graphicx}
\usepackage{dcolumn}
\usepackage{bm}

\newcommand{\be}{\begin{eqnarray}}
\newcommand{\ee}{\end{eqnarray}}
\newcommand{\bse}{\begin{subequations}}
\newcommand{\ese}{\end{subequations}}
\newcommand{\bit}{\begin{itemize}}
\newcommand{\eit}{\end{itemize}}
\newcommand{\ben}{\begin{enumerate}}
\newcommand{\een}{\end{enumerate}}
\newcommand{\bpm}{\begin{pmatrix}}
\newcommand{\epm}{\end{pmatrix}}

\newcommand{\mcal}{\mathcal}

\newcommand{\mrm}{\mathrm}

\newcommand{\bs}{\boldsymbol}

\newcommand{\p}{\partial}
\newcommand{\f}{\frac}
\newcommand{\diff}{\mrm{d}}
\newcommand{\lan}{\left\langle}
\newcommand{\ran}{\right\rangle}

\newcommand{\kB}{k_\mrm{B}}
\newcommand{\eps}{\epsilon}

\begin{document}
\title{\textbf{\textrm{Heat transport and diffusion in a canonical model of a relativistic gas}}}

\author{Malihe Ghodrat}
    \email{ghodrat@shirazu.ac.ir}

\author{Afshin Montakhab}
    \email{montakhab@shirazu.ac.ir}

\affiliation{Department of Physics, College of Sciences, Shiraz
    University, Shiraz 71454, Iran}

\date{\today}

\begin{abstract}
Relativistic transport phenomena are important from both theoretical and practical point of view. Accordingly, hydrodynamics of relativistic gas has been extensively studied theoretically. Here, we introduce a three-dimensional canonical model of hard-sphere relativistic gas which allows us to impose appropriate temperature gradient along a given direction maintaining the system in a non-equilibrium steady state.  We use such a numerical laboratory to study the appropriateness of the so-called first order (Chapman-Enskog) relativistic hydrodynamics by calculating various transport coefficients. Our numerical results are consistent with predictions of such a theory for a wide range of temperatures. Our results are somewhat surprising since such linear theories are not consistent with the fundamental assumption of the special theory of relativity ($v\leq c$). We therefore seek to explain such results by studying the appropriateness of diffusive transport in the relativistic gas, comparing our results with that of a classical gas. We find that the relativistic correction (constraint) in the hydrodynamic limit amounts to small negligible corrections, thus indicating the validity of the linear approximation in near equilibrium transport phenomena.
\end{abstract}


\maketitle
\section{Introduction}
Owing to its diverse applications in the hydrodynamic description of
astrophysical phenomena\cite{Mis68,Dor69,Wei71,And07}, heavy ion
collision experiments  and hot plasmas generated in colliders
\cite{Cse94,Elz01,Xu05,Bir08,Mur10}, relativistic kinetic theory has
attracted considerable attention ever since it was formulated.
The extension of the theory to non-equilibrium situations also covers different problems ranging from formation of the universe to engineering applications such as
pulsed laser cutting and welding or high speed machining processes \cite{Bat02, Ali05}. Perhaps the central issue in this field is the relation between thermodynamic forces and fluxes in a system out of equilibrium. From a phenomenological point of view, in hydrodynamic regime and not too far from equilibrium, fluxes are linearly related to thermodynamic
forces with the so called \emph{transport coefficients} as
proportionality factors. The theories that assume such linear
relations are known as \emph{first-order} theories and are based
on the hypothesis of local equilibrium in the hydrodynamic regime. A
familiar example is the Fourier law of heat transport that relates
heat current to the temperature gradient. The objection raised against such linear laws is that they lead to diffusion equations implying an unbounded propagation of fluctuations in sharp contrast to the first assumption of the special theory of relativity. Various suggestions, including Israel-Stewart second-order
formulation~\cite{Isr79,Ste79,Den10,Bet11} and several more extensions~\cite{Rom10,And10,Lop10} are proposed in the
literature to remove the non-causality and instabilities
associated with first order theories. The price to be paid,
however, is the introduction of complicated higher-order terms
into the formulations as also noted in \cite{Gcol06,Gper09}. It is not our goal here to add a new proposal or even judge the existing theories. Here, we rather propose to study how well (or badly) do first-order theories capture the general features of relativistic transport phenomena.
\par
Our approach is motivated by the need to find the simplest approximate model given both the significance of such problems and the complicated existing alternatives, mentioned above. We therefore propose and study a canonical three-dimensional (3d) model of a relativistic hard-sphere gas in contact with pre-defined temperatures. We are therefore able to impose temperature gradients on our system and use such a realistic numerical laboratory to calculate various transport coefficients in the intermediate and highly relativistic regimes. We then compare our results with the theoretical values obtained from Chapman-Enskog (CE) approximation to relativistic transport equations which constitutes a well-known example of a linear first order theory. We find good agreement between our simulation results and theoretical predictions.  We then seek to explain such agreements by studying diffusive transport in classical versus relativistic dynamics.  In Section \ref{s:Rel-kin-theory}, the main features of
relativistic kinetic theory as well as CE
approximation are reviewed.
Section \ref{s:the model} describes the details of our numerical
model in order to simulate a relativistic hard-sphere gas in contact with a
heat bath. In Section \ref{s:results}, the results of our simulations
and a comparison with previous analytical calculations are
presented. Finally, we close with concluding remarks in Section \ref{s:Conclusion}

\section{Relativistic kinetic theory}\label{s:Rel-kin-theory}
In a statistical description of many-body systems, the
one-particle distribution function is a key ingredient. It can be
interpreted as giving the average number of particles with a
certain momentum at each space-time point and its explicit
expression is obtained by solving a kinetic equation called
transport equation \cite{Hua65,Bal75}. This equation which was
first derived by Boltzmann for a classical gaseous system, gives
the rate of change of the distribution function in time and space
due to the particle interactions. Generally speaking, the form of
transport equation is based upon three assumptions: (i) only binary
collision are considered, (ii) the spatiotemporal changes of
distribution function on microscopic scales are negligible and
(iii) the ``assumption of molecular chaos" (or ``Stosszahlansatz''
\cite{CMax65}) holds. The last one is a
statistical assumption which says no correlations exists between
the colliding particles; that is, the average number of binary
collisions is proportional to the product of their distribution
function. It is therefore evident how the microscopic structures
of the system are reflected in the distribution function and why
it can relate microscopic features to equilibrium and
non-equilibrium macroscopic properties.

\par The relativistic counterpart of the classical transport equation is obtained following
the same reasoning, with same assumptions as
above, but within the framework of relativistic theory
\cite{Gro80,Lib90,Cer02}. \be \label{e:rel_kinetic_equation} p^\mu
\p_\mu f(x,p)=\mcal C[f(x,p)] \ee in which $f(x,p)$ is the
distribution function, $x$ and $p$ are respectively position and
momentum four-vectors, and $\p_\mu=\p/\p
x_\mu=(c^{-1}\p_t,\nabla)$. The collision function, which includes
the particle interactions has the form \be \nonumber \mcal
C[f]&=&\f{1}{2}\int \f{\diff^3 p_1}{p_1^0}\;\f{\diff^3
p'}{p^{'0}}\;\f{\diff^3 p'_1}{p_1^{'0}}\times\\ &&[f'f'_1
W(p',p'_1|p,p_1)-ff_1 W(p,p_1|p',p'_1)],\ee
where $W$ denotes the transition rate and is determined by the
underlying dynamics \cite{Gro80}. The abbreviations $f$ and $f_1$
are used for $f(x,p)$ and $f(x,p_1)$ and the primes refer to
quantities after collision. The solution of the homogeneous
transport equation, i.e. $\mcal C[f]=0$, leads to the equilibrium
distribution, \be \label{e:euilib_dist}f^{(eq)}=\f{\exp[-\beta_\mu p^\mu]}{(2\pi\hbar)^3},\ee in which $(2\pi\hbar)$ denotes Planck constant
and $\beta_\mu=\f{1}{\kB T}U^\mu$ is the temperature four-vector. The hydrodynamic velocity four-vector, $U^\mu=\gamma(u)(c,\bs u)$ with  $\gamma(u)=\sqrt{1-(u/c)^2}$ as Lorentz factor, is defined so that $U_\mu U^\mu=c^2$. Eq.~\eqref{e:euilib_dist} which is best known as J\"uttner distribution, is the relativistic generalization of the Maxwell-Boltzmann distribution and its
validity has been a source of long lasting debates in the physics
community \cite{Lan66,*Lan67,*Kam68,*Kam69,*Lan96,Hor81,*Leh06,*Dun07,Cub07,Nak09,Mon09,Gho09}.
Having this distribution in hand, the particle four-flow in equilibrium is given by \be \label{e:par four-flow} J^\mu_n=\f{c}{(2\pi\hbar)^3}\int \f{\diff^3 p}{p^0}
p^\mu \exp[-\beta_\mu p^\mu],\ee and the equilibrium particle density is defined as $n:=c^{-2}J^\mu_n U_\mu$ \cite{Gro80}. Similarly, the equilibrium energy-momentum tensor can be defined as \be \label{e:energy-mom tensor} \mcal T^{\mu\nu}= \f{c}{(2\pi\hbar)^3}\int \f{\diff^3 p}{p^0}p^\mu p^\nu \exp[-\beta_\mu p^\mu], \ee and the energy density
as $n\epsilon:=c^{-2}\mcal T^{\mu\nu}U_\mu U_\nu$.

\par When the system is not too far from equilibrium, the
transport equation may be solved assuming the distribution
function is a function of local hydrodynamic variables, e.g.
particle density, energy density and hydrodynamic velocity, as
well as their gradients. The systematic method to obtain such
solutions was worked out independently by Chapman \cite{Chap16}
and Enskog \cite{Ens17}. Their approach was based on an expansion of
distribution function around local equilibrium distribution, $f^{(0)}$, with
the ratio of mean free path and a typical macroscopic length as an
expansion parameter, here denoted by $\eta$: \be
\label{e:expansion} f=f^{(0)}+\eta f^{(1)}+ \eta^2 f^{(2)}+\ldots
.\ee The main interest of CE method lies in the first
correction term, which leads to linear laws for the transport
phenomena. Of course, this approximation can only be considered
accurate in the hydrodynamic regime where the expansion parameter
is small.

\par The classical CE approximation can be used in relativistic kinetic theory without major modifications \cite{Sas58,Isr63}.
In the first approximation the one-particle distribution is
written as: \be f(x,p)=f^{(0)}(x,p)[1+\Phi(x,p)],\ee where
$f^{(0)}(x,p)$ is the local relativistic distribution function.
Following the standard kinetic theory, substitution of the
appropriate form of $\Phi$, as a function of hydrodynamic
variables and their gradients, into the relativistic heat flux
expression would give the relativistic generalization of Fourier's
law \cite{Gro80,Gper11}, \be
\label{e:rel_heat_flux}J^\mu_q=\lambda\Delta^{\mu\nu}(\nabla_\nu
T-\f{T}{nh}\nabla_\nu P),\ee in which, $P$ is pressure, $\epsilon$
and $h=\epsilon+Pn^{-1}$, are energy and enthalpy per particle,
and $\Delta^{\mu\nu}:=g^{\mu\nu}-c^{-2}U^\mu U^\nu$ is the
projection operator with $g^{\mu\nu}=\mrm{diag}(1,-1,-1,-1)$ the
metric tensor. It can be shown that the heat transport coefficient
to first order, in CE approximation for
a three dimensional hard-sphere gas is \cite{Gro80,Gper11} \be \label{e:lambda}
\lambda=\f{3}{32\pi}\f{c\kB}{\sigma}\f{z^2
K_2^2(z)[\bar{\gamma}/(\bar{\gamma}-1)]^2}{(z^2+2)K_2(2z)+5zK_3(2z)}\;,\ee
where $z=\f{mc^2}{\kB T}$ and $K_n$'s are the Bessel functions of
the second kind \cite{Abr72}, $\bar{\gamma}=c_P/c_V$, and $\sigma$ is the differential cross section. Using the equation of state of an ideal relativistic gas, $P=n\kB T$, to convert pressure to particle density and
defining $\nabla^\mu$ as $\Delta^{\mu\nu}\partial_\nu$,
Eq.\eqref{e:rel_heat_flux} can be written in the familiar form \be
\label{e:rel_heat_flux_familiar}J^\mu_q=[L_{TT}\f{\nabla^\mu
T}{T}-L_{Tn}\f{\nabla^\mu n}{n}\;], \ee where one defines
\bse\label{e:Onsager_coefficients}\be
L_{TT}&\equiv&\lambda\;T (1-\f{T}{h})\;,\\
L_{Tn}&\equiv&\lambda\; \f{T^2}{h}\;\ee \ese to respectively
distinguish the contributions of energy flux and particle flux in the
total heat flow \cite{Gper11}.

\section{Definition of the model}\label{s:the model}
In \cite{Mon09} we introduced a two dimensional model of a
relativistic hard-disk gas which can be used as a numerical
laboratory to investigate various properties of a relativistic
gas in the low density regime \cite{Mon09,Gho09,Gho11}. Here, we use a three
dimensional generalization of the previous model in which the
particles are spheres rather than disks. The interaction cross-section
associated with hard-disk (or sphere in 3d) model is independent of the energy
and of the scattering angle. This makes it a suitable choice to simulate
high energy hadrons since experimental evidences show that the
total cross sections for massive hadron scattering are more or
less constant in the energy range of interest in relativistic
kinetic theory \cite{Gro80}. For instance, the total cross-section for nucleon-nucleon scattering is roughly constant from an energy of 2GeV on \cite{noteP4_2}. This is the typical energy for the hadron era of the big bang theory \cite{Gro80}.
\par The system we study here, consists of $N$ particles of equal rest mass $m$
which are constrained to move in a box of volume
$V=L_xL_yL_z$. The spherical
particles move in straight lines at constant speed and change
their momenta instantaneously when they touch at distance
$\delta$. The binary collision of particles in three dimension
involves six degrees of freedom (three degrees for each particle)
which can be solved deterministically if we limit ourselves to
head-to-head collisions. In such collisions, the force is exerted
along the line connecting the centers, $\bs r_{ij}=\bs r_i-\bs
r_j$, and thus the two components of particles' momenta
perpendicular to $\bs r_{ij}$ remain unchanged. The parallel components, however, change according to energy-momentum
conservation laws as below \cite{Cub07,Mon09}: \be\label{e:dynamics}
\nonumber p'_{\,i,\parallel}&=&\gamma(v_{cm})^{2}[2v_{cm}\epsilon_{i}-(1+v_{cm}^{2})p_{\,i,\parallel}],\\
\epsilon'_{i}&=&\gamma(v_{cm})^{2}[(1+v_{cm}^{2})\epsilon_{i}-2v_{cm}p_{i,\parallel}],\ee
where hatted quantities refer to momenta after collision and
$v_{cm}=(p\,_{i,\parallel}+p_{j,\parallel})/(E_{i}+E_{j})$ is the
collision invariant, relativistic center-of-mass velocity of the
two particles. With the same rules for particle $j$, a deterministic,
time-reversible canonical transformation at each collision is
defined.

\par A hard-sphere model such as above avoids the
general problems associated with interacting relativistic particle
systems\cite{Whe49}. Thus it can be successfully applied to
simulate a relativistic gas to shed light on challenging questions
like how thermostatistical properties change when the gas is
viewed by a moving observer as compared to a rest observer
\cite{Mon09} or what happens if the measurements are performed
with respect to laboratory time \cite{Mon09}, proper time
\cite{Gho09} or on the light-cone \cite{DunNat09}. These issues
were studied in microcanonical ensemble where any type of
interaction with environment is absent. In this article we
introduce thermal interactions and/or temperature gradients to the
system  to perform simulations in the canonical ensemble or to study
out-of-equilibrium properties.

\par The heat bath we have used here, is a stochastic thermo-bath \cite{Gho11,Teh98, Hat99} which interacts locally, i.e. through the walls, with the particles inside the box. Effectively, it is a random number generator that simulates the exchange of energy between an ideal relativistic gas and the canonical relativistic system at a predefined temperature $T$. Whenever a particle impinges on the boundaries, the
components of its velocity are updated. More specifically, we
remove the incident particle and introduce a new particle with a
new random energy value $\epsilon'$ with the following rates,
while keeping its position unchanged:
\begin{eqnarray}
  \nonumber \mcal W(\eps(v)\rightarrow\eps'(v'))&=&1\; \qquad \;\;\;\mrm{for}\qquad \mcal P_{v'}>\mcal P_{v}\\
  \mcal W(\eps(v)\rightarrow\eps'(v'))&=&\frac{\mcal P_{v'}}{\mcal P_{v}}
  \qquad \mathrm{for}\qquad\mcal P_{v'}<\mcal P_{v}
\end{eqnarray}
This choice of rates is consistent with the condition of detailed
balance \cite{Bel04}. In order to get the correct velocity
distribution, the probability is defined, according to J\"uttner
distribution, as
\be \mcal P_{v}\propto \gamma(v)^{d+2}\exp[-\beta m \gamma(v)], \ee in which
$v$ represents the magnitude of the velocity, $d$ is the space
dimension ($d=3$ here), and the temperature of the bath is introduced
through the parameter $\beta=1/\kB T$ \cite{Mon09,Gho11}. Such thermalization process
allows the particle's energy to be exchanged if the new energy is
more probable ($\mcal P_{v'}>\mcal P_{v}$) while it gives the
system a chance to deviate from the desired distribution ($\mcal
P_{v'}<\mcal P_{v}$). Accepting energies with reduced probability
allows the system to sample fluctuations away from the lowest
energy configuration and thus gives the correct J\"uttner
distribution.

\par In our numerical simulations, the walls normal to the $x$-direction are
diathermal and other walls are periodic. Therefore, the sign of
the $x$-component of the updated velocity of the particle is
manually chosen to make sure that the incident particle reflects back into
the box. Also, the components of the updated velocity should be chosen appropriately so that its magnitude respects the main assumption of relativity, $|v|<c$.

\par Numerical results indicate that the effective temperature obtained using the above
stochastic bath does not exactly meet the nominal value of bath's
temperature, especially when the walls are kept in different
temperatures. This observation which is also reported in other
non-relativistic transport models \cite{Ris96,Lep97} would not
affect our general results; i.e., the existence of a well-defined
stationary state, linear temperature profile and the validity of
local thermal equilibrium \cite{Gho11}.

\section{Results}\label{s:results}
In the following sections the results of our simulations in canonical ensemble and out-of-equilibrium states are presented. In all simulations, the values of rest mass $m$,  particles' diameter $\delta$, Boltzmann constant $\kB$, and light velocity $c$, are all set to unity. To satisfy the condition of zero net momentum in the rest frame, half the particles are given random initial velocities and the second half are chosen to have the same velocities with opposite signs as the first half. The total momentum remains zero (on the average) after the system is allowed to interact with the heat baths and thus the assumption of zero hydrodynamic velocity is valid. To obtain the velocity distributions, we let the system equilibrate (typically after $10^{2}N$ collisions) and simultaneously measure velocities of all particles with respect to
a rest frame at a given instant of time $t$. To collect more data we may either repeat the
above procedure for different initial conditions or perform
measurements at several well-separated time instances.

\subsection{Canonical ensemble}\label{s:results-a}
\begin{figure}
  \centering
  \includegraphics[width=9cm]{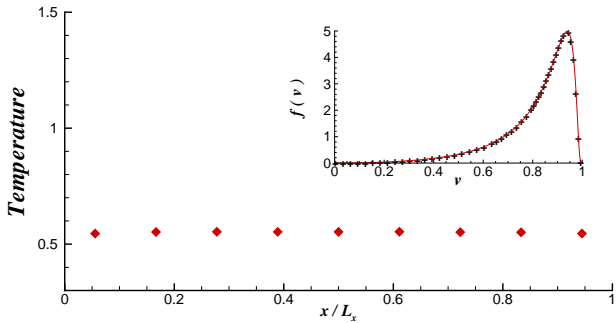}\\
  \caption{\label{f:canonical_ensemble}(color online). Constant temperature profile $T(x)$ for a canonical simulation. $N=256$
  particles were put in a box of size $L_xL_yL_z$ with its diathermal walls kept at the same temperature.
  The system is divided into equally-spaced vertical bins and the temperature for each bin has
  been calculated using Eq.\eqref{Juttner-energy}. Inset shows
  the measured velocity distribution (+) of particles
  which is well fitted to the J\"uttner function (solid red line) with the expected temperature.}
\end{figure}
The model described above lead to a well-defined
equilibrium state in the canonical ensemble. To show this we have measured the
velocity distribution of the particles with respect to a rest
observer and laboratory time $t$ \cite{noteP4}. The result shown
in Fig.\ref{f:canonical_ensemble} clearly indicates that the
particles have reached a state consistent with the prediction of
equilibrium relativistic statistics, i.e., the J\"uttner velocity
distribution and a constant temperature profile. To obtain the
temperature profile we typically divide our system into $\mrm{M}$
equal bins of volume $L_xL_yL_z/M$. Each bin is large enough to
allow computation of (local) relevant thermodynamic variable. For
instance, the average number of particles in each bin is computed
by
\begin{equation}
n_{k}=\int_{\Delta x}\frac{1}{t}\int_{0}^{t}\sum_{i=1}^{N}
\delta(x-x_{i}(t'))dt'dx,
\end{equation}
in which, $x_{i}(t')$ is the position of $i$-th particle at time
$t'$ and $\Delta x$ is the region that contains position of the $k^{th}$ bin, i.e.,
$[x_k-\frac{\Delta x}{2},x_k+\frac{\Delta x}{2}]$. The same
technique is used to obtain the average energy, $\varepsilon_{k}$,
in each bin:
\begin{equation} \varepsilon_{k}=\int_{\Delta
x}\frac{1}{t}\int_{0}^{t}\sum_{i=1}^{N}
\delta(x-x_{i}(t'))\epsilon_i(t')dt'dx,
\end{equation}
The temperature profile, $T_{k}$, is then computed as a function
of $n_{k}$ and $\varepsilon_{k}$ using the following equation
\cite{Mon09,Han09}: \be \label{Juttner-energy}
\frac{\varepsilon_{k}}{n_{k}}=\frac{3}{\beta_k}+m\;\frac{K_1(m\beta_k
)}{K_2(m\beta_k)\;},\ee which is written for $d=3$.

\subsection{Out-of-equilibrium studies}\label{s:results-b}
\begin{figure}
  \centering
  \includegraphics[width=8.5cm]{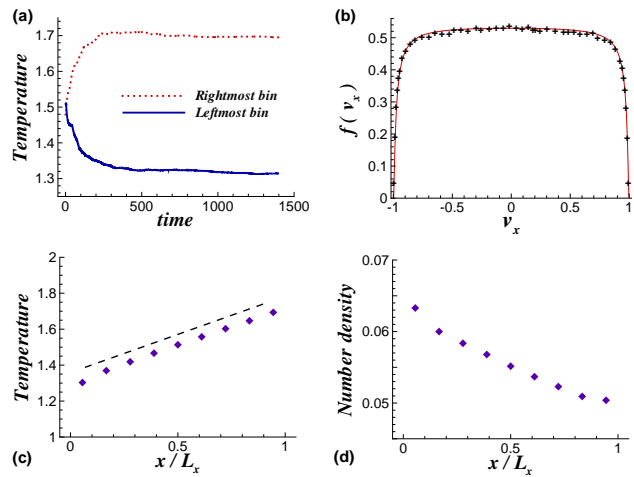}\\
  \caption{\label{f:temp_density_prof}(color online). (a) Shows changes in temperature of leftmost and rightmost
  bins immediately after the system is exposed to a temperature gradient. In the
  resulting non-equilibrium stationary state, the velocity distribution of the middle bin of
  the system is well fitted to the J\"uttner function with the expected
  temperature. The linear behavior of temperature profile (c) and the
  resulting gradient in particle density (d) are also shown.}
\end{figure}

We now turn to the out of equilibrium situation in which the
vertical left and right walls are kept at different temperatures.
Results of simulation with $N=256$ particles and box of size $L_x=45,\; L_y=L_z=10$ indicate that
after typically $10^{2}N$ collisions, a non-equilibrium stationary
state is obtained. In this state the average value of
thermodynamic properties remain constant and the condition of
local thermal equilibrium is satisfied (see
Fig.\ref{f:temp_density_prof}). That is, for small temperature
gradients, in every macroscopically infinitesimal volume element,
thermodynamic variables can be defined in the usual way and are
related to each other through the relations which hold for the
system at equilibrium. To show this, we have divided the system
into nine bins. The temperature profile, $T_{k}$, is computed as
discussed in the previous section.
Fig.\ref{f:temp_density_prof}(a) shows how the temperature of
different bins (here the leftmost and rightmost bins) change after the introduction of different baths until a non-equilibrium stationary state is
achieved. In this state a \textit{linear} temperature profile is obtained (Fig.\ref{f:temp_density_prof}(c)) which
indicates that $\nabla T(x)$ is constant throughout the system.
The temperature gradient imposed, on the other hand, causes a
gradient in particle density (Fig.\ref{f:temp_density_prof}(d)) as
expected from Onsager relations \cite{Ons31}. Additionally, the
velocity distribution for a typical bin (see
Fig.\ref{f:temp_density_prof}(b)) is well fitted to J\"uttner
distribution with the corresponding temperature from
Fig.\ref{f:temp_density_prof}(c). These conditions indicate that the obtained stationary state is locally equilibrated and thus can be used to study transport
properties of relativistic ideal gases.

\subsection{Heat transport coefficients}\label{s:results-c}

\begin{figure}
  \centering
  \includegraphics[width=9cm]{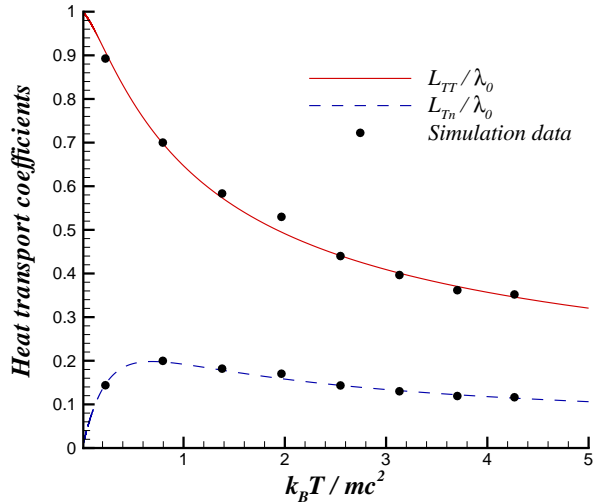}\\
  \caption{\label{f:Onsager_coef}(color online). The numerical (dots) and theoretical (lines)
  heat transport coefficients of a relativistic gas as a function of temperature.
  The coefficients $L_{TT}$ (upper branch) and $L_{Tn}$ (lower branch), are normalized to the low temperature limit $\lambda_0=75 \kB\sqrt{\kB T/m\pi}/(128\sigma)$.
  The nominal values of temperature difference in
  all simulations is $\Delta T=0.5$. Note that in the relativistic high temperature regime
  the contribution of particle flux is comparable to the contribution of
  energy flow as predicted by the relativistic theory of irreversible processes to linear order.}
\end{figure}

Our main concern in this section is to study the heat transport
properties of the hard-sphere relativistic gas. We therefore propose to verify the validity of linear approximation (i.e. Eq~\eqref{e:rel_heat_flux_familiar}), by directly calculating the heat flux in the presence of a temperature gradient and thus calculating the relevant transport coefficients and comparing them with theoretical prediction Eq.~\eqref{e:Onsager_coefficients}. In order to numerically obtain the value of $\lambda$ in this
equation, we impose a fixed temperature gradient, as
described above, and directly measure the
energy flow in our simulations using the following formula: \be
\label{e:rel-energy-flow-2}\bs J_\epsilon=\f{1}{V}\lan \sum_{r=1}^{N}
\epsilon_r \bs v_{r} \ran, \ee where $\lan\ldots\ran$ denotes time
averaging. Note that the system is simulated in the rest frame,
i.e the average total momentum of the system in all directions and thus the average hydrodynamic
velocity is zero. Therefore, the origin of energy flow is purely
chaotic, that is $\bs J_\epsilon=\bs J_q$ \cite{Gper12}. Since the temperature gradient is imposed in the $x$
direction the only nonzero component of heat current is
$J_q^{(x)}$. Other numerically measured quantities are $\nabla T$,
$\nabla n$, $T$, and $h$. As stated in Sec. \ref{s:the model}, the
temperature of walls do not (exactly) coincides with the nominal
temperature of the heat baths. Therefore, the effective
temperature gradient obtained in simulations, $\nabla T^\ast$, as
well as numerical value of $\nabla n$ is substituted in
Eqs.~\eqref{e:rel_heat_flux_familiar}. The values of temperature, $T$,
and enthalpy, $h$, are mean values evaluated for the middle bin.
Due to linear behavior of temperature profile, the temperature of
the middle bin can also be expressed as the average of the
corresponding values in the rightmost and leftmost bins. It is now
straightforward to calculate heat transport coefficient,
$\lambda$, and consequently the coefficients $L_{TT}$ and
$L_{Tn}$, defined in Eq.\eqref{e:Onsager_coefficients}.
Figure~\ref{f:Onsager_coef} shows the results obtained by averaging over $10^6$ realizations. The coefficients
are normalized to the low temperature limit of Eq.~\eqref{e:lambda},
$\lambda_0=\f{75\kB}{128\sigma}\sqrt{\f{\kB T}{m\pi}}$ \cite{Gro80}. As indicated,
the behavior of the numerically obtained transport coefficients
agree well with theoretical predictions of relativistic
CE approximation, (Eqs.\eqref{e:lambda} and \eqref{e:Onsager_coefficients}).
Our results therefore show the validity of such linear approximation for a wide
range of temperatures well into the relativistic regime.

\par It is worth emphasizing that the dynamics of the hard-sphere model
we have used here fully respects the assumptions of relativity, in the
sense that the velocity of particles as well as propagation speed
of fluctuations in thermodynamic properties such as energy, heat,
density, etc. are all less than the maximum allowed velocity, $c$.
Under such circumstances, the agreement between numerical results
and theoretical predictions of relativistic CE
approximation indicates that this linear theory is capable of
obtaining heat transport coefficients of a relativistic gas. The
validity of CE approximation, however, is limited by the ratio of
mean free path to the characteristic length scale (or relaxation
time scale to the dynamical time scale) of the system which forms
the expansion parameter $\eta$ in Eq.\eqref{e:expansion}. This
parameter should be small enough to let us ignore nonlinear terms.
It is therefore expected that CE fails when superfluid helium,
rarefied gasses, or nanoscale problems are studied. This
classically accepted criterion also holds in relativistic regime. However, it is not straightforward to numerically investigate the validity of higher order terms via the present (low density) model. In this model we have neglected issues such as the possibility of radius contraction of moving spheres, the possibility of multiple (non-binary) collisions, and the importance of particle interactions beyond hard sphere potential. We are not aware of a deterministic MD model concerning all theses issues though in some kinetic models, Boltzmann transport equation has been solved for a dense ultra-relativistic gas considering binary (and including) three-body interactions \cite{Xu05,Xu13}.

\par Another important result is the contribution of particle flow in the flux of heat
(see Fig.\ref{f:Onsager_coef}.). As is well-known, heat
flow is due to the diffusive transport of energy in a
non-relativistic limit and the contribution of particle flow,
$L_{Tn}$, in the flux of heat is well negligible. However, as
can be seen in Fig.\ref{f:Onsager_coef}, the contribution of particle flow
becomes an increasingly significant part of heat flux as one
moves towards the relativistic, high temperature regime. This purely
relativistic effect add more subtlety to the theory and thus could lead to interesting nontrivial consequences when other thermodynamic are considered (See below for more discussions.).

\subsection{Diffusive transport of heat}\label{s:results-d}
\begin{figure}
  \centering
  \includegraphics[width=8.5cm]{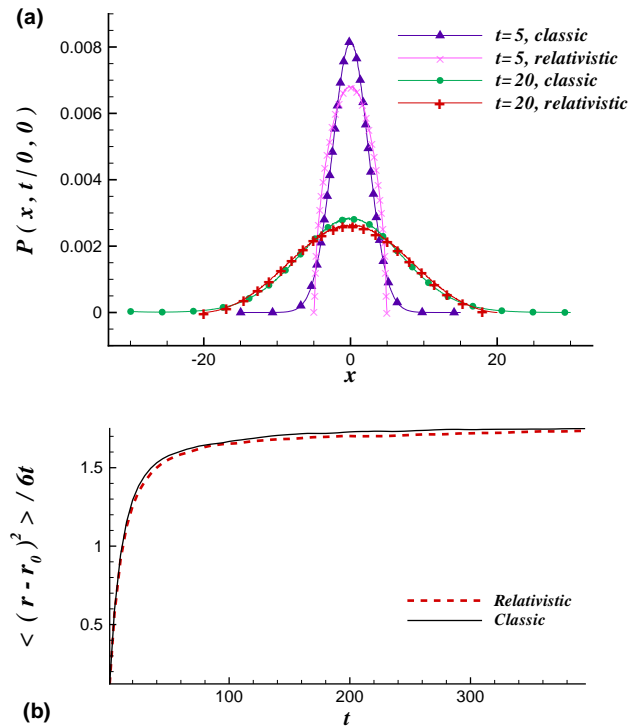}\\
  \caption{\label{f:propagators} (color online). Part (a) shows short time and long time behavior of relativistic propagators obtained for a $3d$ gaseous system with particles of diameter $\delta=1$, number density $n=0.064$. Both systems are initiated with same initial conditions. The corresponding mean free time and temperature of classical and relativistic gas are computed to be $MFT_{cl.}=3.57$, $MFT_{rel.}=3.55$, $T_{cl.}=0.33$ and $T_{rel.}=2.66$. Part (b) shows the diffusion coefficient as a function of time that fits to the curve $D(t)=1.85-\f{2.09}{\sqrt{t}}$ for classical data and agrees well with the analytically predicted value $\mcal D=1.89$. Note that the relativistic coefficient is always less than the classical value due to the constraints of relativity.}
\end{figure}
In Section \ref{s:results-c} we showed that the heat flux in a relativistic system is well described by the linear approximation Eq.~\eqref{e:rel_heat_flux_familiar}. Since no net flux of particles is present in the non-equilibrium stationary state (i.e. $J^\mu_n=L_{nT}\f{\nabla^\mu
T}{T}-L_{nn}\f{\nabla^\mu n}{n}=0$), it is possible to write Eq.~\eqref{e:rel_heat_flux_familiar} in the closed form \be J^\mu_q=-\kappa \nabla^\mu T\ee
in which $\kappa=(L_{nT}^2-L_{TT}L_{nn})/(T L_{nn})$ is the thermal conductivity. Combining this linear law with the conservation of energy-momentum tensor ($\p_\mu \mcal T^{\mu\nu}=0$) and using the relation $\mcal T^{00}=c_V T$, which is valid for both low and high temperature limits, results in the following  diffusion equation for propagation of heat: \be \label{e:T_diff}\f{\p T(\bs r,t)}{\p t}=\mcal D_T \nabla^2 T(\bs r,t),\ee where $\mcal D_T=-\kappa/c_V$, $\mcal T^{\mu\nu}$ denotes energy-momentum tensor and $c_V$ is the molar heat capacity. On the other hand, if the system is put between two particle reservoirs with different chemical potentials, the same type of equation is obtained for density: \be \label{e:N_diff}\f{\p n(\bs r,t)}{\p t}=\mcal D_n \nabla^2 n(\bs r, t)\ee

\par The diffusive propagation of heat/density is known to be inconsistent with the basic assumption of relativity as it allows superluminal transport. The underlying mechanism for such macroscopic relaxation process is the random walk nature of Brownian particles also described by a diffusion equation whose solution is given by propagator in the $d$-dimensional space, \be p(\bs r,t|\bs r_0,0)=\f{1}{(4\pi \mcal D t)^{d/2}}\exp(\f{-(\bs r-\bs r_0)^2}{4 \mcal D t}),\ee where $\mcal D$ is the appropriate diffusion constant, and $\bs r_0$ is the position of the Brownian particle at $t=0$. The propagator give the probability of finding the particle at time $t$ in position $\bs r$.  The nonzero probability at arbitrary long distances for short times is the basis of the objection to such diffusive properties within relativistic theories.  While such an objection is clearly justified, the question we ask is that under typical relativistic parameters how differently does a Brownian particle move as compared to its classical counterpart.  We therefore calculate such propagators directly by monitoring a ``tagged" particle. In order to simulate identical situations, the initial conditions (i.e. initial positions and velocities) are chosen to be the same in both classical and relativistic systems. Fig.~\ref{f:propagators}(a) shows the propagators in the $x$-direction. As expected, the classical propagator fits to a Gaussian  which is basically different from relativistic propagator in the sense that the latter is restricted to the region $|\bs x- \bs x_0|<c(t-t_0)$ while the former allows an unbounded propagation of particles (or heat). This fundamental difference is the origin of objections against diffusive transport of particles (or heat) and consequently against the so called first order theories. Given such discrepancy, the question is why the relativistic first order theory of CE is good enough when diffusion or heat transport coefficients of a relativistic system are considered? The reason traces back to the hydrodynamic limit in which the long time (compared to mean free time) behavior of the system is considered.

\par As Fig.~\ref{f:propagators}(a) indicates, in this limit the difference of classical and relativistic propagators diminishes so that the classical Gaussian propagator (namely, the solution of diffusion equation) becomes an almost exact approximation of the relativistic propagator. Consequently, the resulting relativistic diffusion coefficient shows no major distinction from its classical value as indicated in Fig.~\ref{f:propagators}(b). In this figure the classical as well as relativistic diffusion coefficient are plotted as a function of time. The transient behavior seen between ballistic ($<(\Delta \bs r)^2> \propto t^2$) and diffusive ($<(\Delta \bs r)^2> \propto t$) regimes is a sign of correlations ( or memory effects) in the system which is well studied in the literature \cite{Rei80}. It can be shown that the diffusion coefficient of a classical hard-sphere gas fits to the curve $ D(t)=\mcal D-\f{cte.}{\sqrt{t}}$ which, after sufficiently long time, saturates to the constant coefficient predicted for a simple uncorrelated model (i.e., $\mcal D=3/(8n\delta^2)\sqrt{\kB T/m\pi}$ with $n$ denoting number density \cite{Rei80}.). Also, note that the value of relativistic diffusion is always less than the classical one due to the well known relativistic constraints. The important point to be noted here is the similarity of classical and relativistic coefficients which clearly supports the appropriateness of first order theories and the consequent diffusion equation in the hydrodynamic limit.

\par Here, we would like to make an important point. Despite the fact that we have argued that diffusion equation is a fairly good approximation in the hydrodynamic limit, there is still a fundamental difference between the classical and relativistic propagators in Fig~\ref{f:propagators}. Surely, they look very similar, but nevertheless one is a Gaussian and the other is \emph{not}. The non-Gaussian nature of the relativistic propagator is due to the relativistic constraint $|v|<c$ which effectively introduces some sort of correlation/memory in the random walk process thus making it a non-markovian process \cite{Hak68,Han09}. Such correlations may have important consequences as far as cooperative phenomena are considered. If such phenomena exist we expect them to become more pronounced in the high temperature (relativistic) regime as such correlations become more pronounced. This is potentially very interesting since cooperative phenomena in non-interacting systems are rare and thus unexpected results. A closer study of this issue is out of the scope of this manuscript and deserves a separate publication \cite{Gho12}.

\section{Concluding remarks} \label{s:Conclusion}
In this article we have presented results of numerical simulation of a three-dimensional canonical generalization of our previous model of a relativistic gas. We have used such a numerical laboratory in order to study transport properties in the presence of a temperature gradient. We use parameter regimes where local equilibrium is shown to be well-established. Our main result is that various transport coefficients are consistent with the CE (linear) approximation to the relativistic transport equation. We observe such consistency for a wide range of temperature well into the high relativistic regime. Our results are somewhat surprising because of the well-known inconsistency between first order theories and the fundamental assumption of special relativity. We have therefore proposed to look into transport propagators both for the classical as well as highly relativistic gas in order to find an explanation for the observed accuracy of the linear approximation. We find that in the long time hydrodynamic limit, the kinetic description of such diffusive processes are almost identical, with practically negligible difference between the classical and the relativistic regime. The Gaussian nature of such propagators and the resulting diffusion coefficient is much the same as the relativistic one despite the fact that the latter respects the constraints of special relativity. Our results provide a basis for justification of applicability of first order relativistic transport theories despite their well-known limitations.
\bibliographystyle{apsrev}
\bibliography{xbib}

\end{document}